\documentclass[graybox]{svmult}

\usepackage{wrapfig}
\usepackage{helvet}         % selects Helvetica as sans-serif font
\usepackage{courier}        % selects Courier as typewriter font
\usepackage{type1cm}   
\usepackage{amssymb}
\usepackage{makeidx}         % allows index generation
\usepackage{graphicx}        % standard LaTeX graphics tool
\usepackage{multicol}        % used for the two-column index
\usepackage[bottom]{footmisc}% places footnotes at page bottom

\def\be{\begin{equation}}
\def\ee{\end{equation}}    

\def\ba{\begin{eqnarray}}
\def\ea{\end{eqnarray}}    
 
\def\p{\partial}

\makeindex             % used for the subject index
\begin{document}

\title*{Hamilton--Jacobi Method and Gravitation}
\author{R. Di Criscienzo, L. Vanzo and S. Zerbini}
\institute{R. Di Criscienzo \at Dipartimento di Fisica, Universit\`{a} di Trento and INFN; \email{rdicris@science.unitn.it}
\and L. Vanzo \at Dipartimento di Fisica, Universit\`{a} di Trento and INFN; \email{vanzo@science.unitn.it}
\and S. Zerbini \at Dipartimento di Fisica, Universit\`{a} di Trento and INFN; \email{zerbini@science.unitn.it}}
\maketitle
% 
% \abstract*{Each chapter should be preceded by an abstract (10--15 lines long) that summarizes the content. The abstract will appear \textit{online} at \url{www.SpringerLink.com} and be available with unrestricted access. This allows unregistered users to read the abstract as a teaser for the complete chapter. As a general rule the abstracts will not appear in the printed version of your book unless it is the style of your particular book or that of the series to which your book belongs.
% Please use the 'starred' version of the new Springer \texttt{abstract} command for typesetting the text of the online abstracts (cf. source file of this chapter template \texttt{abstract}) and include them with the source files of your manuscript. Use the plain \texttt{abstract} command if the abstract is also to appear in the printed version of the book.}
% 
\abstract{Studying the behaviour of a quantum field in a classical, curved, spacetime is an extraordinary task which nobody is able to take on at present time. Independently by the fact that such problem is not likely to be solved soon, still we possess the instruments to perform exact predictions in special, highly symmetric, conditions. Aim of the present contribution is to show how it is possible to extract quantitative information about a variety of physical phenomena in very general situations by virtue of the so-called Hamilton--Jacobi method. In particular, we shall prove the agreement of such semi-classical method with exact results of quantum field theoretic calculations.}

\section{Introduction}
\label{sec:1}

Suppose we are interested in studying the behaviour of a field $\Phi(x)$ (scalar, for sake of simplicity) in a curved spacetime endowed with a (trapping) horizon (e.g. in the vicinity of a black hole). Based on physical intuition, we expect that the interaction from the quantum field and the classical background gives rise to different phenomena, as: Hawking radiation through the horizon; decay of unstable particles scattering off the gravitational field; vacuum particle creation in regions of strong gravity; radiation from (possibly, naked) singularities, etc. The aformentioned topics would pertain the investigation of a quantum theory of gravity, the lack of which obliges us to work with standard techniques.\\
The field is governed by the Klein--Gordon equation,
\be
\left(\Box_x  - \frac{m^2}{\hbar^2}\right) \Phi(x) = 0 \;,\label{KG}
\ee
where the parameter $m^2$ is interpreted as the field mass and for convenience we have inserted $\hbar$ explicitly.
Field quantization can be performed in either canonical or path integral ways:
\be
\Phi(\mathbf{x}) = \int_{\mathcal{A}(\mathbf{x})} \mbox{D}p \cdot \mbox{D}\tilde{\mathbf{x}} \cdot\mbox{D}N \times \left(\mbox{Gauge-fixing conditions}\right)\times \exp(i\, I[p,\tilde{\mathbf x},N]) \label{pi}
\ee
where $\mathcal{A}(\mathbf{x})$ represents appropriate boundary conditions and $I[p,x,N]$ is the Hamiltonian formulation of the action. The fact that $\Phi(N,\mathbf{x})$ has to satisfy the equation of motion (\ref{KG}) imposes constraints on the allowed boundary conditions, $\mathcal{A}$.\\ 
Solution to (\ref{pi}) is largely unknown due to the difficulty of computing path integrals in curved spacetimes. However, some information is accessible in the WKB regime of approximation. In this case, one generally finds appropriate to look for a solution in the form,
\be
\Phi(x) = D(x) e^{-  \frac{I(x)}{\hbar}} + O(\hbar) \label{exp}
\ee
where the small parameter $\hbar$ is used to govern the WKB expansion. Inserting (\ref{exp}) into (\ref{KG}) and equating powers of $\hbar$, we obtain to the lowest orders
\ba
\hbar^{-2} &:& \qquad -\nabla^a I \nabla_a I + m^2 =0  \nonumber \\
\hbar^{-1} &:& \qquad 2 \nabla D \cdot \nabla I + D \Box I =0 \nonumber \\
\hbar^{0}  &:& \qquad \Box D = 0 \nonumber \\
\cdots &:& \qquad \cdots \label{HJ1}
\ea 
Exact computation of the pre-factor $D(x)$ is complicated even in very addomesticated situations and therefore it is beyond our present goal. \\
Let us split $I$ into a real and a purely imaginary part: $I(x) := I_R(x) - i S(x)$; then (\ref{HJ1}) becomes:
\be
- (\nabla I_R)^2 + (\nabla S)^2 + m^2 = 0. \label{HJ2}
\ee
If the imaginary part of $I$ varies with $x$ much more rapidly than the real part, that is, if $\vert \nabla S\vert \gg \vert \nabla I_R\vert $, it follows from  (\ref{HJ2}) that $S$ will be an approximate solution to the (Lorentzian) Hamilton-Jacobi equation
\be
g^{a b} \p S_a \p S_b + m^2 = 0 \label{HJ3} .
\ee 
Furthmore, the wave function (\ref{exp}) will then be predominantly of the form $e^{i S}$. Of course, going from the exact path integral form (\ref{pi}) to the approximate regime (\ref{exp}) with $I$ solution to (\ref{HJ1}), we loose specification of the boundary conditions $\mathcal{A}$. It will be evident later that this lost is only apparent.

The basic idea proposed some time ago by Parick \& Wilzcek \cite{pw} is to interpret the spacetime horizon -- say, for example, of a black hole -- as a sort of barrier  and to study the tunnelling of field quanta through it. Certainly, the  horizon behavies in quite a different way with respect to usual quantum mechanical potential barriers. In ordinary quantum mechanics, the barrier is represented by the region between the turning points of the classical trajectories. Here instead, the horizon is just a point on the classical characteristic curves. Evaluating the tunnelling in quantum mechanics means computing the ratio between particle wavefunction on the two sides of the barrier. In the black hole case, instead, the required ratio is generated by a discontinuity in the wavefunction. Moreover, in the familiar context of tunneling through a barrier, an imaginary part comes from a negative eigenvalue of the small disturbance operator around the classical bounce, while the Euclidean action is real. In the black hole case, instead, as we shall see later, it is the action itself that is complex. 

Given the whole sort of specifications above, we can conclude the reasoning and invoke the well known result according to which, the creation probability per unit time of quanta of mass $m$ is given -- to leading order in $\hbar$ -- by the WKB formula
 \begin{equation}
\Gamma \propto \exp \left(- \frac{2}{\hbar}S\right)
\label{prob}
\end{equation}
with $S$ solution to (\ref{HJ3}). Remarkably, as it has been shown in \cite{anvz,nvz,km,dnvzz,hdnvz}, the procedure outlined so far reproduces the infamous Planckian spectrum of Hawking radiation in the case of black hole spacetimes: $\Gamma \propto \exp(-\beta\omega_H)$, with $\omega_H$ the energy of tunnelling particles through the black hole horizon and $\beta$ interpreted as the inverse temperature of the thermalized field quanta. The Hamilton--Jacobi method of tunnelling has therefore proved an elegant way to interpret Hawking radiation as a tunnelling process and to derive in relatively simple way the associated temperature ($T = \beta^{-1}$). As we shall try to show in the following, the method does not exhaust its power in the computation of black hole Hawking temperature, generalizing indeed to a wider class of spacetime horizons (e.g. cosmological horizons) and to other kinds of semi-classical phenomena.  

We use the conventions according to which the metric signature is $(-,+,+,+)$; first latin indices as $a,b$ run over $0,\dots,3$, mid-latin indices as $i,j$ only over $0,1$. From nown on, we implement natural units, so that $c= \hbar = G = k_B =1$.

\section{The Kodama--Hayward formalism for spherically symmetric spacetimes}
\label{sec:2}

In the following, we shall limit ourselves to focus only on spherically symmetric spacetimes where no gravitational waves production is involved. The line element can be locally written as \cite{hay}
\be
ds^2 = \gamma_{ij}(x) dx^i dx^j + R^2(x) d\Omega^2 \;, \label{metric}
\ee
where the two-dimensional metric $\gamma_{ij}(x)$ is referred to as the normal metric, $\{x^i\}$ are associated coordinates and $R(x^i)$ is the areal radius, considered as a scalar field in the two-dimensional normal space. We recall that to have a truly dynamical solution, i.e. to avoid Birkhoff's theorem, the spacetime must be filled with matter everywhere. Examples are the Vaidya solution, which contains a flux of radiation at infinity, and FRW solutions which contain a perfect fluid.\\
A dynamical trapping horizon, if it exists, is located at $0 = \chi(x)\vert_H$, with $\chi(x):= \gamma^{ij} (x)\p_i R(x)\p_j R(x)$, provided that $\p_i \chi(x)= 0$. The dynamical surface gravity associated with the horizon is given by the normal space scalar $\kappa_H = \frac{1}{2} \Box_{\gamma} R(x) \vert_H$ as proved in \cite{hay}. \\
In the spherical symmetric dynamical case, it is possible to introduce the so-called Kodama vector field $K$, with $(K^aG_{a b})^{;b} = 0$, that can be taken as its defining property, \cite{kod}. It follows that $K$ is a natural generalization of the Killing vector of stationary spacetimes. Given the metric (\ref{metric}), the non-vanishing Kodama vector components are $K^i= \epsilon^{ij}\p_j R(x)/\sqrt{-\gamma}$ ($\epsilon^{0 1}=+1$). The Kodama vector gives a preferred flow of time and in this sense it generalizes the flow
of time given by the Killing vector in the static case. As a consequence, we may introduce the invariant energy associated with a particle of mass $m$ by means of the scalar quantity on the normal space 
\be
\omega= - K\cdot d S \label{energy}
\ee 
where $S$ is the particle action which we assume to satisfy the reduced Hamilton--Jacobi equation 
\be
\gamma^{ij} \p_i S \p_j S + m^2 =0\; .\label{rhj}
\ee

Remarkably, the probability rate (\ref{prob}) does not depend by the choice of coordinates, since the horizon location, the horizon surface gravity, the Kodama energy are all invariantly defined in the space normal to the spheres of symmetry \cite{dhnvz}.

The basic idea can now be roughly described as follows: the reduced Hamilton--Jacobi equation (\ref{rhj}) supplemented by the Kodama energy formula (\ref{energy}) constrains particle's momenta, e.g. $\p_+ S = \p_+ S(x^{\pm}, m,\omega)$; thus, the mass parameter $m$ gives two complementary energy scales so that, according to the physical phenomenon involved, the two scales exchange the leading role in the analysis.\\
More in detail, suppose we are interested in the physics of the horizon: tunnelling through the horizon -- typically related to Hawking/Unruh effects -- corresponds to the existence of a simple pole in particle's momenta. In this case, it turns out that the mass parameter can be neglected so that, to all the extents, particles move along null trajectories. On the other hand, if we are now interested in bulk effects, away from any horizon, then the mass parameter plays a crucial role being possibly responsable for a branch point singularity in tunnelling particle's momenta.\\
Let us make an example in order to make clearer what we mean. As fully described in \cite{dvz}, the FRW spacetime with spatial curvature $\hat k = \frac{k}{l^2}$ ($k=0,\pm 1$ and $l$ an opportune length scale) represents a dynamical, spherically symmetric spacetime exhibiting a cosmological horizon in correspondence of what we shall call the Hubble radius, namely $R_H (t):= (H^2+ \hat k/a^2)^{-1/2}$ and $R(t,r):=a(t) r$. The Kodama energy is $\omega = \sqrt{1-\hat k r^2}(-\partial_t I + r H \partial_r I)$ $\equiv \sqrt{1-\hat k r^2}\,\tilde{\omega}$. The Hamilton--Jacobi equation reads $-(\p_t S)^2 + \frac{(1-\hat k r^2) }{a^2(t)} \,(\p_r S)^2 + m^2=0$, so that the radial particle's momentum is
\be
\p_r S=-\frac{ a H \tilde\omega (ar) \pm a\sqrt{\omega^2 -m^2(1-(a r/R_H)^2)}}{1-(a r/R_H)^2}\;.
\label{g}
\ee
Near the horizon, the mass coefficient vanishes so that we can set $m=0$. Thus, making a null-horizon radial expansion, the action for particles coming out of the horizon towards the inner (untrapped) region is $S = 2 \int dr \p_r I$, with $\p_r S$ exhibiting a simple pole at the horizon. To deal with the simple pole in the integrand, we implement Feynman's $i\varepsilon${--}prescription, something which resambles the recovering of the boundary conditions encoded in the path integral approach mentioned above. Because of (\ref{prob}), $\Gamma \sim \exp(-\omega_H/T)$, $\omega_H >0$ for physical particles and $T=-\kappa_H/2\pi$ ($\kappa_H <0$ for trapping horizons of the inner type such as the Hubble radius, cf. \cite{hay2} ) the dynamical temperature associated to FRW horizon.\\
To treat instead the decay of unstable composite particles inside the Hubble horizon (i.e., in the untrapped region), we need to identify the energy of the particle before the decay as the Kodama energy, $\omega$; then we denote by $m$ the effective mass parameter of one of the decay products, after the decay. With these understandings, we find out that for the unstable particle sitting at rest at the origin of the comoving coordinates, there is an imaginary part of the action as the decay product tunnels into the region $0<r <r_0$ to escape beyond $r_0$, with $r_0$ implicitly defined through $[a(t) r_0]^2 = R_0^2 := \left(1- \frac{\omega^2}{m^2}\right) R_H^2$. Assuming a two-particle decay, the rate is
\be
\Gamma=\Gamma_0\,e^{-2\pi\,R_H\,(m-\omega)}   
\label{g1}
\ee
and $\Gamma_0$ depending on the interaction coupling (e.g. $\Gamma_0\sim\lambda^2$ for a $\lambda\phi^3$ interaction). Equation (\ref{g1}) agrees with Volovik result for de Sitter space \cite{vol} and with asymptotic quantum field theory calculation by Bros \textit{et al}, \cite{bem}.

\section{Vacuum Particle Creation and Emission from Naked Singularities}
\label{sec:3}

A perfectly legitimate question we can ask ourselves is weather the method is extendable to the case of static black holes as well. With regard to this, we consider the exterior region of a spherically symmetric, static, black hole spacetime and repeat the same argument. Quite generally, we can write the line element as
\be
ds^2= -e^{2\psi(r)} C(r) dt^2 + \frac{dr^2}{C(r)} + r^2 d \Omega^2 \;.
\ee
The analysis of the radial momentum is made easier by setting the Kodama energy $\omega =0$: in the intention, this would correspond to particle creation from vacuum,
\be
\int dr\,\p_r S =  m \int_{r_1}^{r_2} dr \frac{1}{\sqrt{-C(r)}}\;.
\label{vacuum}
\ee
The integration is taken over any interval $(r_1,r_2)$ where $C(r) >0$. Equation (\ref{vacuum}) shows that, under very general conditions, in static black hole spacetimes there could be a decay rate whenever a region where $C(r)$ is positive exists. However, it is an easy task to show that the spacelike singularity of the Schwarzschild black hole does not emit particles in the semi-classical regime: in the interior, the Kodama vector is spacelike, thus no energy can be introduced there.\\
The situation is very different when a naked singularity is present. Considering a neutral particle
in the Reissner-Nordstr\"{o}m solution with mass $M$ and charge $Q > 0$ (for definiteness), the line element is
\be
ds^2 = - \frac{(r-r_-)(r-r_+)}{r^2} dt^2 + \frac{r^2}{(r-r_-)(r-r_+)} dr^2 + r^2 d\Omega^2\;,
\ee
with $r_\pm$ functions of $(M,Q)$ denoting the inner and outer horizons. The function $C(r)= (r-r_-)(r-r_+)/r^2$ is negative in between the two horizons, where the Kodama vector is spacelike, so there the action is real. On the other hand, it is positive within the outer communication domain, $r > r_+$, but also within the region contained by the inner Cauchy horizon, that is $0 < r < r_-$. Thus, because of (\ref{vacuum}) and assuming the particles come created in pairs, we obtain that, modulo the pre-factor over which we have nothing to say, there is a creation probability per unit time and unit volume (equation (\ref{rn}) not depending upon the creation event) of neutral particles of mass $m$ by the strong gravitational field near the Reissner-Nordstr\"{o}m naked singularity $(M,Q)$ which goes as
\be
\Gamma \sim \left(\frac{M-Q}{M+Q}\right)^{mM}e^{-2Qm}.\label{rn}
\ee
At a first look, the process of particle production in the region close to the singularity raises the issue of the stability of the solution. However, this does not seem to be a problem. In fact, radiation created by the bulk close to the singularity comes into the singularity with infinite red-shift and approaches the future inner, classically unstable, horizon with infinite blue-shift. Thus, the contribution of the radiation coming in the singularity to the back-reaction is negligible and the causal structure of the singularity safe; while the blue-shifted radiation approaching the future sheet of the inner horizon will contribute to its quantum instability (Cf. \cite{fn} for further investigation).

A complementary and potentially interesting effect is the emission from the naked singularity itself. We investigate this problem for the case of two-dimensional dilaton gravity, and will come back to Reissner-Nordstr\"{o}m solution afterward.\\
Consider the two-dimensional metric
\be
ds^2=\sigma^{-1}dx^+dx^- \;,\qquad\sigma:=\lambda^2x^+x^--a(x^+-x_0^+)\theta(x^+-x_0^+)
\ee
where $\lambda$ is related to the cosmological constant by $\Lambda=-4\lambda^2$ and $a$ represents the wave amplitude. 
This metric arises as a solution of $2D$ dilaton gravity coupled to a bosonic field with stress tensor $T_{++}=2a\,\delta(x^+-x_0^+)$, describing a shock wave. $\sigma=0$ is a naked singularity partly to the future of a flat space region (linear dilaton vacuum). The heavy arrow in the figure represents the history of the shock wave responsible for the existence of the timelike singularity.
\begin{figure}[h]
\sidecaption
\includegraphics[scale=.40]{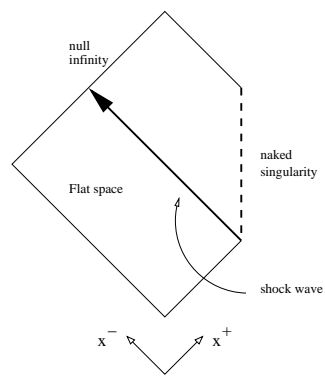}
\caption{The naked singularity formed by the shock wave.}
\label{fig:1}       % Give a unique label
\end{figure}
The goal is to compute the outgoing flux which is given by: $2T_{++}-2T_{--}$. In order to do this, we notice that the Hamilton--Jacobi equation implies either $\p_+ S = 0$ or $\p_-S = 0$, $S$ being the action. To find the ingoing flux we
integrate along $x^+$ till we encounter the naked singularity, using $\p_-S = 0$, so that $S=\int dx^+\partial_+I = \int dx^+\frac{\omega}{\sigma}$. The absorption probability is $\Gamma(\omega)=\Gamma_0\,\exp[-2\pi\omega/(\lambda^2 x^- -a)]$. The flux is computed by integrating the probability over the coordinate frequency $\tilde{\omega}=\frac{\omega}{2\sigma}$ (that is, the variable conjugate to the time coordinate), with the density of states measure $\frac{d\tilde{\omega}}{2\pi}$: $T_{++}=\Gamma_0\frac{(\lambda^2x^- -a)^2}{2\pi^3\sigma^2}$.\\
To find $T_{--}$, we integrate now along $x^-$ starting from the naked singularity, this time
using $\partial_+ S=0$. A similar calculation gives $T_{--}=\Gamma_0\frac{\lambda^4 (x^+)^2}{2\pi^3\sigma^2}$. \\
$T_{+-}$ is given by the conformal anomaly, $T=4\sigma T_{+-}=R/24 \pi$ (for one bosonic degree of freedom). Matching to the anomaly gives $\Gamma_0=\pi^2/24 \sim O(1)$.\\
Note that the stress tensor diverges approaching the singularity, indicating that its resolution will not
be possible within classical gravity but requires instead quantum gravity \cite{giapp}.\\
Indeed, all these results agree with the one-loop calculation to be found in \cite{vw}.

Returning now to the Reissner-Nordstr\"{o}m solution, could it be that the naked singularity emits
particles? In the four-dimensional case one easily sees that the action has no imaginary part along null
trajectories either ending or beginning at the singularity. Formally this is because the
Kodama energy coincides with the Killing energy in such a static manifold and there is no
infinite red-shift from the singularity to infinity. Even considering the metric as a genuinely two-dimensional, however, it is possible to show that the action does not exhibit any imaginary part \cite{dvz}. \\
It is fair to say that the  Reissner-Nordstr\"{o}m naked singularity will not emit particles in
this approximation something which seems to be coherent with quantum field theory results, \cite{dvz, fp}.

\section{Conclusions}
\label{sec:4}

We have shown that semi-classical tunnelling method can handle several quantum effects: radiation from dynamical horizons (both cosmological and collapsing); gravitational enhancement of particle decay otherwise forbidden by
conservation laws; radiation from two-dimensional naked singularities. Normally, great efforts are needed to analyze quantum effects in gravity, while instead the tunneling picture promptly gives strong indications of what's going on. The obtained agreement between both the particle decay rates and the radiation from naked singularities in the tunnelling picture
and the (asymptotic of the) exact results -- when they exist in particular conditions -- gives confidence, in our opinion, of the
validity of the method even in more general situations.

\end{document}